\documentclass[12pt, a4paper]{article}
\usepackage{amsfonts,amssymb,amsmath,amsthm,graphicx}

\begin{document}

\title{Casimir interaction of two plates inside a cylinder.}

\author{Valery N.Marachevsky \thanks{email: maraval@mail.ru} \\
{\it V. A. Fock Institute of Physics, St. Petersburg
State University,}\\
{\it 198504 St. Petersburg, Russia} }



\maketitle

\begin{abstract}
The new exact formulas for the attractive Casimir force acting on
each of the two identical perfectly conducting plates moving
freely inside an infinite perfectly conducting cylinder with the
same cross section are derived at zero and finite temperatures by
making use of the zeta function technique. The long and short
distance behaviour of the plates' free energy is investigated.
\end{abstract}

\maketitle

\section{Introduction}

Recently a new geometry in the Casimir effect\cite{Casimir}, a
piston geometry, has been introduced  in a $2D$ Dirichlet model
\cite{Cavalcanti}. Generally the piston is located in a
semi-infinite cylinder closed at its head. The piston is
perpendicular to the walls of the cylinder and can move freely
inside it. The cross sections of the piston and cylinder coincide.
Physically this means that the approximation is valid when the
distance between the piston and the walls of a cylinder is small
in comparison with the piston size.

In paper \cite{Jaffe2} a perfectly conducting square piston at zero
temperature was investigated in $3D$ model in the electromagnetic
and scalar case. The exact formula (Eq.(6) in \cite{Jaffe2}) for the
force on a square piston was written in the electromagnetic case.
Also the limit of short distances was found for arbitrary cross
sections at zero temperature (Eq.(7) in \cite{Jaffe2}).

A dilute circular piston and cylinder were studied perturbatively in
\cite{Barton}. In this case the force on two plates inside a
cylinder and the force in a piston geometry differ essentially. The
force in a piston geometry can even change sign in this
approximation for thin enough walls of the material. Other examples
of pistons in a scalar case were presented in \cite{Fulling, Edery,
Zhai}.

In this paper we consider  a $3D$ model in the electromagnetic case
with physically realistic perfectly conducting boundary conditions.
Two identical plates move freely inside an infinite cylinder with
the same cross section, which is arbitrary. The plates are
perpendicular to the walls of the cylinder. In Section $2$ we derive
the new exact zero temperature result (\ref{r7}), (\ref{p10}) for
the Casimir energy of two identical parallel plates inside a
cylinder with an arbitrary cross section by making use of the zeta
function technique \cite{Santangelo2, Elizalde}. Two special cases
of (\ref{p10}), the exact results for rectangular and circular
cylinders, are briefly discussed. Also we discuss the important
property of the result (\ref{p10}) - its exponential damping at long
distances (\ref{om1}). In Section $3$ the new exact result
(\ref{r21}) for the free energy of two identical parallel plates
inside an infinite cylinder is derived. In the long distance limit
the new high temperature result (\ref{r24}) for the free energy is
obtained. In Section $4$ the heat kernel technique
\cite{Vassilevich, Gil, Marach1} is applied to derive the leading
short distance behaviour of the free energy. Also we prove that in
the short distance limit of the high temperature result (\ref{r24})
the known high temperature result for two infinite perfectly
conducting parallel plates (\ref{om2}) is obtained.

We take $\hbar=c=1$.

\section{Zero temperature results}

Our aim is to calculate the Casimir energy of interaction and the
force between the two identical parallel plates of an arbitrary
cross section inside an infinite cylinder of the same cross
section (the plates are perpendicular to the walls of the
cylinder).

$TE$ and $TM$ eigenfrequencies of the perfectly conducting
cylindrical resonator with an arbitrary cross section $M$ can be
written as follows. For $TE$ modes ($E_z =0 $) inside the
perfectly conducting cylindrical resonator $[0,a]\times M$ with
$n$ being a normal to the boundary $\partial M$ of an arbitrary
cross section $M$ the magnetic field $B_z (x,y,z)$ and
eigenfrequencies $\omega_{ TE}$ are determined by:
\begin{align}
& B_{z} (x,y,z) = \sum_{i=1, l=1}^{+\infty} B_{il}
\sin \Bigl(\frac{\pi l z}{a}\Bigr) g_i (x,y) , \\
& \Delta^{(2)} g_i (x,y) = - \lambda_{i N}^2 g_i (x,y) \\
& \frac{\partial g_i (x,y)}{\partial n} \Bigl|_{\partial M} = 0 \\
& \omega_{TE}^2 =  \Bigl(\frac{\pi l}{a}\Bigr)^2 + \lambda_{i N}^2
, \:\: , \lambda_{i N}\ne 0,  \:\: l,i = 1 \: .. +\infty .
\end{align}
The other components of the magnetic and electric fields can be
expressed via $B_z (x,y,z)$.

For the $TM$ modes ($B_z =0$) inside the same perfectly conducting
cylindrical resonator the electric field $E_z (x,y,z)$ and
eigenfrequencies $\omega_{ TM}$ are determined by:
\begin{align}
& E_{z} (x,y,z) = \sum_{l=0, k=1}^{+\infty} E_{kl}
\cos \Bigl(\frac{\pi l z}{a}\Bigr) f_k(x,y) , \\
& \Delta^{(2)} f_k (x,y) = - \lambda_{k D}^2 f_k (x,y) \\
&  f_k (x,y) |_{\partial M} = 0 \\
& \omega_{TM}^2 =  \Bigl(\frac{\pi l}{a}\Bigr)^2 + \lambda_{k D}^2
, \:\: l=0 \: .. +\infty , k=1 \: .. +\infty
\end{align}

In  zeta function regularization scheme the Casimir energy is
defined as follows:
\begin{equation}
E = \frac{1}{2} \Bigl(\sum\omega_{TE}^{1-s} +
\sum\omega_{TM}^{1-s} \Bigr) \Bigl|_{s=0} , \label{a2}
\end{equation}
where the sum has to be calculated for large positive values of
$s$ and after that an analytical continuation to the value $s=0$
is performed.

Alternatively one can define the Casimir energy via a zero
temperature one loop effective action $W$ ($T_1$ is a time
interval here):
\begin{align}
&W = - E T_1  \\
&E = - \zeta^{\prime} (0) \label{a3} \\ &\zeta(s) =
\frac{1}{\Gamma(\frac{s}{2})} \int_{0}^{+\infty} dt\,
t^{\frac{s}{2} - 1} \nonumber \\ &\sum_{\omega_{TE},\omega_{TM}}
\int_{-\infty}^{+\infty} \frac{d p}{2\pi} \exp \biggl( - t
\Bigl(\frac{a}{\pi}\Bigr)^2 \Bigl(\omega^2 +p^2\Bigr)\biggr)
\label{a1}
\end{align}
After integration over $p$ in (\ref{a1}) one can see that
definitions (\ref{a2}) and (\ref{a3}) coincide.

In every Casimir sum it is convenient to write:
\begin{equation}
\sum_{l=1}^{+\infty} \exp(- t l^2) = \frac{1}{2} \, \theta_3
\Bigl(0, \frac{t}{\pi}\Bigr) - \frac{1}{2} . \label{a4}
\end{equation}
For the first term on the right-hand side of (\ref{a4}) we use the
property of the theta function $\theta_3(0,x)$:
\begin{equation}
\theta_3 (0, x) = \frac{1}{\sqrt{x}} \, \theta_3
\Bigl(0,\frac{1}{x}\Bigr)
\end{equation}
and the value of the integral
\begin{equation}
\int_0^{+\infty} dt \, t^{\alpha-1} \exp \Bigl(-p\: t
-\frac{q}{t}\Bigr) = 2 \Bigl(\frac{q}{p}\Bigr)^{\frac{\alpha}{2}}
K_{\alpha} (2 \sqrt{p q} )
\end{equation}
for nonzero values of $n$ to rewrite the Neumann zeta function
$\zeta_N (s)$ (arising from TE modes) in the form:
\begin{multline}
\zeta_N (s) = \\ = \sum_{\lambda_{iN}} \int_{-\infty}^{+\infty}
\frac{dp}{2\pi} \biggl[
\frac{\sqrt{\pi}\:\Gamma\bigl((s-1)/2\bigr)}{2 \:\Gamma(s/2)}
\Bigl(\frac{a\sqrt{\lambda_{iN}^2 + p^2}}{\pi}\Bigr)^{1-s}
 + \\ + \sum_{l=1}^{+\infty} \frac{2\sqrt{\pi}}{\Gamma(s/2)}
\Bigl(\frac{\pi^2 l}{a \sqrt{\lambda_{iN}^2
+p^2}}\Bigr)^{\frac{s-1}{2}} K_{\frac{s-1}{2}} \Bigl(2al
\sqrt{\lambda_{iN}^2 +p^2}\Bigr) \biggr] -  \\ -
\sum_{\lambda_{iN}}\frac{\sqrt{\pi}\:\Gamma\bigl((s-1)/2\bigr)}{4
a \Gamma(s/2)} \Bigl(\frac{a \lambda_{iN}}{\pi}\Bigr)^{1-s}
\label{a5}
\end{multline}

The Neumann part of the Casimir energy is given by:
\begin{multline}
E_N = \sum_{\lambda_{iN}} \int_{-\infty}^{+\infty} \frac{d p}{2
\pi} \frac{1}{2} \ln \Bigl(1-
\exp(-2 a \sqrt{\lambda_{iN}^2 +p^2}) \Bigr) +  \\
+ \frac{a}{2} \sum_{\lambda_{iN}} \int_{-\infty}^{+\infty} \frac{d
p}{2 \pi}  \Bigl( \lambda_{iN}^2 +p^2
\Bigr)^{\frac{1-s}{2}}\biggr|_{s=0} - \frac{1}{4}
\sum_{\lambda_{iN}} \lambda_{iN}^{1-s}\biggr|_{s=0} . \label{a6}
\end{multline}
Here we used $K_{-1/2} (x) =\sqrt{\pi/(2\,x)}\exp(-x)$.

The Dirichlet part of the Casimir energy (from TM modes) is
obtained by analogy:
\begin{multline}
E_D = \sum_{\lambda_{kD}} \int_{-\infty}^{+\infty} \frac{d p}{2
\pi} \frac{1}{2} \ln \Bigl(1-
\exp(-2 a \sqrt{\lambda_{kD}^2 +p^2}) \Bigr) +  \\
+ \frac{a}{2} \sum_{\lambda_{kD}} \int_{-\infty}^{+\infty} \frac{d
p}{2 \pi}  \Bigl( \lambda_{kD}^2 +p^2
\Bigr)^{\frac{1-s}{2}}\biggr|_{s=0} + \frac{1}{4}
\sum_{\lambda_{kD}} \lambda_{kD}^{1-s}\biggr|_{s=0}  .\label{a7}
\end{multline}

The electromagnetic Casimir energy of a perfectly conducting
resonator of the length $a$ and an arbitrary cross section is
given by the sum of (\ref{a6}) and (\ref{a7}) :
\begin{align}
E = &\sum_{\lambda_{iN}} \int_{-\infty}^{+\infty} \frac{d p}{2
\pi} \frac{1}{2} \ln \Bigl(1- \exp(-2 a \sqrt{\lambda_{iN}^2
+p^2}) \Bigr) + \label{a8} \\ + &\sum_{\lambda_{kD}}
\int_{-\infty}^{+\infty} \frac{d p}{2 \pi} \frac{1}{2} \ln
\Bigl(1- \exp(-2 a \sqrt{\lambda_{kD}^2
+p^2}) \Bigr) + \label{a9} \\
+ &\frac{a}{2} \sum_{\lambda_{iN}} \int_{-\infty}^{+\infty}
\frac{d p}{2 \pi}  \Bigl( \lambda_{iN}^2 +p^2
\Bigr)^{\frac{1-s}{2}}\biggr|_{s=0} + \label{a10} \\ +
&\frac{a}{2} \sum_{\lambda_{kD}} \int_{-\infty}^{+\infty} \frac{d
p}{2 \pi}
\Bigl( \lambda_{kD}^2 +p^2 \Bigr)^{\frac{1-s}{2}}\biggr|_{s=0} +
\label{a11}\\
+ &\frac{1}{4} \sum_{\lambda_{kD}} \lambda_{kD}^{1-s}\biggr|_{s=0}
- \frac{1}{4} \sum_{\lambda_{iN}} \lambda_{iN}^{1-s}\biggr|_{s=0}
. \label{a12}
\end{align}

The terms
\begin{align}
E_{cylinder} = &\frac{1}{2} \sum_{\lambda_{iN}}
\int_{-\infty}^{+\infty} \frac{d p}{2 \pi}  \Bigl( \lambda_{iN}^2
+p^2 \Bigr)^{\frac{1-s}{2}}\biggr|_{s=0} + \nonumber \\ +
&\frac{1}{2} \sum_{\lambda_{kD}} \int_{-\infty}^{+\infty} \frac{d
p}{2 \pi} \Bigl( \lambda_{kD}^2 +p^2
\Bigr)^{\frac{1-s}{2}}\biggr|_{s=0} \label{om8}
\end{align}
yield the electromagnetic Casimir energy for a unit length of a
perfectly conducting infinite cylinder with the same cross section
$M$ as the resonator under consideration.

For rectangular boxes it was generally believed \cite{Lukosz,
Bordag1}
 that the repulsive contribution to the force acting on two parallel
opposite sides  of a {\it single} box (and resulting here from
(\ref{a10}-\ref{a11})) could be measured in experiment. However, the
terms  (\ref{a10}-\ref{a11}) are closely related to the Casimir
energy of an infinite in $z$ direction cylinder when there are no
plates inside (see eq.(\ref{om8})). Imagine that the box is large in
$z$ direction. Its Casimir energy and the Casimir energy of an
infinite in $z$ direction cylinder coincide when the two opposite
$z$ sides of the box are located at spatial infinity. To calculate
the energy change between these two configurations and the force on
a $z$ side of the box one should subtract the energy of an infinite
cylinder from the expression for $E$ (\ref{a8}-\ref{a12}) when the
box sides are infinitely far from each other. Then the force on a
$z$ side of the box is equal to zero for infinite distance $a$
between box $z$ sides.

For the experimental check of the Casimir energy one should
measure the force somehow. One can insert two parallel perfectly
conducting plates inside an infinite perfectly conducting cylinder
and measure the force acting on one of the plates as it is being
moved through the cylinder.  The distance between the inserted
plates is $a$.


\begin{figure}
\centering \includegraphics[width=10cm]{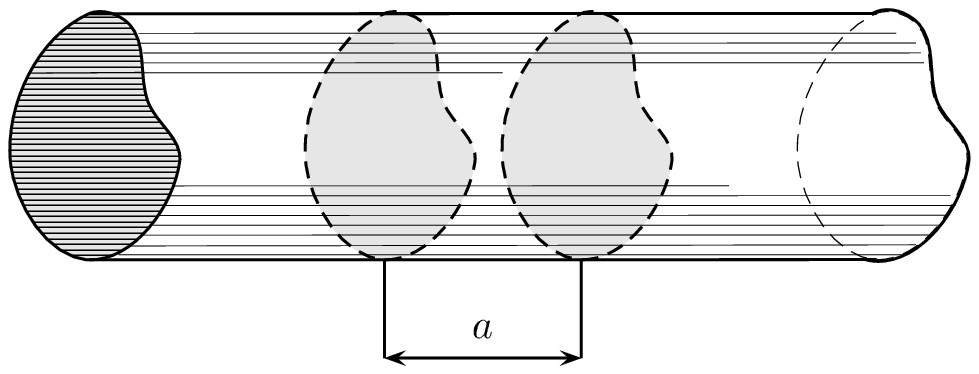} \caption{Two
plates inside an infinite cylinder}
\end{figure}

To calculate the force on each of the two plates inside a cylinder
with the cross section $M$ one can perform the following gedanken
experiment that was frequently used to calculate the Casimir force
between two infinite parallel perfectly conducting plates. Imagine
that $4$ parallel plates are inserted inside an infinite cylinder
and then $2$ exterior plates are moved to spatial infinity. This
situation is exactly equivalent to $3$ perfectly conducting cavities
touching each other. From the energy of this system one has to
subtract the Casimir energy of an infinite cylinder without plates
inside it, only then do we obtain the energy of interaction between
the interior parallel plates, the one that can be measured in the
experiment. Doing so we obtain the attractive force on each interior
plate inside the cylinder:
\begin{align}
F (a) &= - \frac{\partial \mathcal{E}(a)}{\partial a}, \label{a13} \\
\mathcal{E} (a) &= \sum_{\omega_{wave}} \frac{1}{2} \ln (1-\exp(-2 a
\, \omega_{wave})), \label{r7}
\end{align}
the sum here is over all  $TE$ and $TM$ eigenfrequencies
$\omega_{wave}$ for a cylinder with the cross section $M$ and an
infinite length.

In fact, the final results for this geometry should coincide with
the results for the three plates' piston geometry when one of the
three piston plates (the exterior plate) is moved to infinity. Also
one can immediately obtain the result for three plates' system
inside a cylinder, which is exactly the piston geometry, employing
the same arguments and renormalization scheme. In the three plates'
system the force on the interior plate is equal to the sum of the
forces acting on this plate from the two exterior plates, i.e. the
piston geometry can be effectively considered as $2$ two plates'
systems. It should be emphasized that these assertions are valid for
perfect boundary conditions.




By making use of an identity \cite{Nesterenko}
\begin{equation}
\frac{1}{2} \int_{-\infty}^{+\infty} \frac{dp}{2\pi} \ln
\bigl(1-\exp(-2a\sqrt{\lambda^2 + p^2})\bigr) = -
\frac{\lambda}{2\pi} \sum_{l=1}^{+\infty} \frac{K_1(2l\lambda\,
a)}{l}
\end{equation}
 one can rewrite (\ref{r7}) in the form:
 \begin{equation}
 \mathcal{E}(a) = -\frac{1}{2\pi} \sum_{l=1}^{+\infty} \biggl(\sum_{\lambda_{k D}}
 \frac{\lambda_{k D} K_1 (2 l\lambda_{k D} a)}{l}
 + \sum_{\lambda_{i N}}
  \frac{\lambda_{i N} K_1 (2 l\lambda_{i N} a)}{l}\biggr)
  .\label{p10}
 \end{equation}

The results (\ref{r7}) , (\ref{p10}) are our main zero temperature
results. Our results are {\it exact for an arbitrary curved geometry
of a cylinder}.  This fact may be important for the experimental
check of the Casimir effect in piston related geometries with curved
boundaries. One can choose an arbitrary curved plate geometry $M$,
for this geometry the eigenvalues of the two dimensional Dirichlet
and Neumann problems $\lambda_{kD}$, $\lambda_{iN}$ can be found
numerically. After that the exact expressions (\ref{r7}),
(\ref{p10}) can be used to obtain the Casimir force on a plate. In
fact, similar in the form mathematical results can be obtained in
the case of a one dimensional massive scalar field theory on a
manifold with boundary, however, in our case an infinite number of
"masses" $\lambda_{kD}$, $\lambda_{iN}$ appear in the theory due to
existence of the cylinder cross section.



For convenience of the reader we write explicitly two special
cases: rectangular and circular cylinders. For a rectangular
cylinder with the sides $b$ and $c$ the exact Casimir energy of
two plates inside it can be written as:
\begin{multline}
\mathcal{E}_{rect} (a) =
-\sum_{l=1}^{+\infty}\sum_{m,n=-\infty}^{+\infty \: \prime}
\frac{\sqrt{m^2/b^2 + n^2/c^2}}{4l} \:\:  K_{1} (2 l \pi a
\sqrt{m^2/b^2 + n^2/c^2}) . \label{p11}
\end{multline}
The prime means that the term $m=n=0$ is omitted in the sum.

For a circular cylinder the eigenvalues of the two dimensional
Laplace operator $\lambda_{k D} , \lambda_{i N}$ are determined by
the roots of Bessel functions and derivatives of Bessel functions.
The exact Casimir energy of two circular plates of the radius $b$
separated by a distance $a$ inside an infinite circular cylinder
of the radius $b$ is given by:
\begin{align}
&\mathcal{E}_{circ}(a) =  - \sum_{l=1}^{+\infty} \sum_{\nu =
0}^{+\infty} \sum_{j} \frac{1}{2\pi b}\frac{\mu_{D \nu j} K_1 (2 l
\mu_{D \nu j} a/b )
 +  \mu_{N \nu j} K_1 (2 l \mu_{N \nu j} a/b)}{l} ,\label{p12}\\
    &J_{\nu}(\mu_{D \nu j}) = 0 ,
    \:\:\:\: J_{\nu}^{'} (\mu_{N \nu j}) = 0   . \nonumber
 \end{align}
The sum is over positive $\mu_{D \nu j}$ and $\mu_{N \nu j}$.

The leading asymptotic behaviour of $\mathcal{E}(a)$ for long
distances $\lambda_{1 D} a \gg 1$, $\lambda_{1 N} a \gg 1$ is
determined by the lowest positive eigenvalues of the two dimensional
Dirichlet and Neumann problems $\lambda_{1 D}, \lambda_{1 N} $:
\begin{equation}
\mathcal{E}(a)|_{\lambda_{1 D} a \gg 1, \: \lambda_{1 N} a \gg 1 }
\sim - \frac{1}{4\sqrt{\pi a}} \Bigl( \sqrt{\lambda_{1D}} e^{-2
\lambda_{1D} a} + \sqrt{\lambda_{1N}} e^{-2 \lambda_{1N} a} \Bigr) ,
\label{om1}
\end{equation}
so the Casimir force between the two plates in a cylinder is
exponentially small for long distances. This important property of
the solution is due to the gap in the frequency spectrum or, in
other words, it is due to the finite size of the cross section of
the cylinder.

\section{Finite temperature results}
To get the free energy $\mathcal{F} (a,\beta)$ for bosons at nonzero
temperatures ($T=1/\beta$) one has to make the substitutions:
\begin{align}
 p &\to p_m = \frac{2\pi m}{\beta} , \\
 \int_{-\infty}^{+\infty}  \frac{dp}{2\pi} &\to \frac{1}{\beta}
 \sum_{m=-\infty}^{+\infty}.
\end{align}
Thus the free energy describing the interaction of the two
parallel perfectly conducting plates inside an infinite perfectly
conducting cylinder of an arbitrary cross section has the form:
\begin{align}
& \mathcal{F} (a,\beta) = \nonumber \\ & =\frac{1}{\beta}
\sum_{\lambda_{k D}} \sum_{m=-\infty}^{+\infty} \: \frac{1}{2} \ln
\Bigl(1-\exp
\bigl(-2a\sqrt{\lambda_{k D}^2 + p_m^2} \bigr) \Bigr) + \nonumber \\
&+ \frac{1}{\beta} \sum_{\lambda_{i N}} \sum_{m=-\infty}^{+\infty}
\: \frac{1}{2} \ln \Bigl(1-\exp \bigl(-2a\sqrt{\lambda_{i N}^2 +
p_m^2} \bigr) \Bigr) . \label{r21}
\end{align}
This is our central finite temperature result. Note that
$\lambda_{iN} \ne 0$. For rectangular and circular cylinders one
can substitute the explicitly known $\lambda_{kD}$, $\lambda_{iN}$
to (\ref{r21}) in analogy to (\ref{p11}) and (\ref{p12}).

The attractive force between the plates inside an infinite
cylinder of the same cross section at nonzero temperatures is
given by:
\begin{align}
F (a, \beta) &= -\frac{\partial \mathcal{F} (a, \beta) }{\partial a}
= - \frac{1}{\beta} \sum_{\omega_{TD}}
\frac{\omega_{TD}}{\exp(2a\omega_{TD}) - 1} \nonumber
 \\ &  - \frac{1}{\beta}
\sum_{\omega_{TN}} \frac{\omega_{TN}}{\exp(2a\omega_{TN}) - 1}.
 \label{r30}
\end{align}
Here $\omega_{TD} = \sqrt{p_m^2 + \lambda^2_{k D}} $ and
$\omega_{TN} = \sqrt{p_m^2 + \lambda^2_{i N}} $.

In the long distance limit $a \gg \beta/(4\pi)$ one has to keep only
$m=0$ term in $(\ref{r21})$. Thus the free energy of the plates
inside a cylinder in the {\it high temperature limit} is equal to:
\begin{multline}
\mathcal{F}(a, \beta)|_{a \gg \beta/(4\pi)} = \frac{1}{2\beta}
\sum_{\lambda_{k D}}  \: \ln \Bigl(1-\exp (-2a\lambda_{k D} ) \Bigr)
+
\\ +  \frac{1}{2\beta} \sum_{\lambda_{i N}}
\: \ln \Bigl(1-\exp (-2a\lambda_{i N} ) \Bigr) .\label{r24}
\end{multline}
One can check that the limit $\lambda_{1 D} a \ll 1$, $\lambda_{1 N}
a \ll 1$ in (\ref{r24}) immediately yields the known high
temperature result for two parallel perfectly conducting plates
separated by a distance $a$ (see eq.(\ref{om2}) for details).

If the conditions $\lambda_{1 D} a \gg 1$, $\lambda_{1 N} a \gg 1$
are satisfied in addition to the condition $a \gg \beta/(4\pi)$
then the leading asymptotic behaviour of the free energy can be
expressed via the lowest positive eigenvalues of the two
dimensional Dirichlet and Neumann problems $\lambda_{1 D}$ and
$\lambda_{1 N}$ as follows:
\begin{equation}
\mathcal{F}(a,\beta)|_{a \gg \beta/(4\pi) ,\: \lambda_{1 D} a \gg 1,
\: \lambda_{1 N} a \gg 1 } \sim -\frac{1}{2\beta} \bigl(
e^{-2a\lambda_{1D}} + e^{-2a\lambda_{1N}} \bigr),
\end{equation}
so the force between the two plates in a cylinder is exponentially
small for large enough distances between them at finite
temperatures as well.


\section{Free energy: short distance behaviour}

It is convenient to apply the technique of the heat kernel and zeta
function to obtain the short distance behaviour of the free energy
(\ref{r21}). It can be done by noting that if the heat kernel
expansion
\begin{equation}
\sum_{\lambda_i} e^{-t\lambda_i^2} \underset{t\to 0}{\sim}
\sum_{k=0}^{+\infty} t^{\frac{-d+k}{2}} c_k \label{om5}
\end{equation}
exists ($d$ is a dimension of the Riemannian manifold) then one
can write the expansion
\begin{equation}
\sum_{\lambda_i} e^{-\sqrt{t}\lambda_i} \underset{t\to 0}{\sim}
\sum_{k=0}^{d-1} \frac{2 \;\Gamma(d-k)}{\Gamma((d-k)/2)} \;\;
t^{\frac{-d+k}{2}} c_k  \label{r22}
\end{equation}
by making use of the analytical structure of the zeta function.

The proof can be done as follows. Zeta function can be written in
two different forms:
\begin{align}
\zeta(s) &= \sum_{\lambda_i} \frac{1}{\Gamma(s/2)}
\int_{0}^{+\infty} dt \, t^{\frac{s}{2}-1} \exp (-t\lambda_i^2) , \label{p20}\\
\zeta(s) &= \sum_{\lambda_i} \frac{1}{2\Gamma(s)} \int_{0}^{+\infty}
dt \, t^{\frac{s}{2}-1} \exp (-\sqrt{t}\lambda_i) . \label{p21}
\end{align}
It is well known that residues at the poles of the zeta function are
related to the coefficients $c_k$ of the heat kernel expansion
(\ref{om5}):
\begin{equation}
 c_k = \frac{1}{2} \: {\rm Res}_{s=d-k} \bigl(\zeta(s)\Gamma(s/2)\bigr), \label{p22}
\end{equation}
which follows from (\ref{p20}) and (\ref{om5}). The expansion
(\ref{r22}) now follows from (\ref{p21}) and (\ref{p22}).

The free energy (\ref{r21}) can be rewritten as follows:
\begin{multline}
\mathcal{F}(a,\beta) =  \sum_{\lambda_{rD}^{(3)}} \frac{1}{\beta}
\ln \Bigl(1 - \exp (-2a \lambda_{rD}^{(3)} ) \Bigr) +
\sum_{\lambda_{kD}}
\frac{1}{2\beta} \ln \Bigl(1 - \exp (-2a \lambda_{kD} ) \Bigr) + \\
+ \sum_{\lambda_{sN}^{(3)}} \frac{1}{\beta} \ln \Bigl(1 - \exp
(-2a \lambda_{sN}^{(3)} ) \Bigr) - \sum_{\lambda_{iN}}
\frac{1}{2\beta} \ln \Bigl(1 - \exp (-2a \lambda_{iN} ) \Bigr) -
\\ - \sum_{\lambda_{lN}} \frac{1}{\beta}
\ln \Bigl(1 - \exp(-2 a \lambda_{lN}^{(1)})  \Bigr) \label{om4}
\end{multline}
Here the eigenvalues $\lambda_{rD}^{(3)}$ satisfy the equation
$\Delta^{(3)} p_r (x,y,z) = - \lambda_{rD}^{(3)2} p_r (x,y,z) $ and
$p_r(x,y,z)$ satisfy Dirichlet boundary conditions at the boundary
of the manifold $[0,\beta/2]\times M $, the eigenvalues
$\lambda_{sN}^{(3)}$ satisfy the equation $\Delta^{(3)} p_s (x,y,z)
= - \lambda_{sN}^{(3)2} p_s (x,y,z) $ and $p_s(x,y,z)$ satisfy
Neumann boundary conditions at the boundary of the manifold
$[0,\beta/2]\times M $, $M$ is an arbitrary cross section of the
cylinder. Here we sum over all nonzero eigenvalues
$\lambda_{sN}^{(3)}$.

The eigenvalues $\lambda_{lN}^{(1)}$ satisfy the one dimensional
Laplace equation with Neumann boundary conditions at the boundary
of the manifold $[0,\beta/2]$. They appear due to the condition
$\lambda_{iN} \ne 0$ in equation (\ref{r21}) and our decision to
sum over all nonzero $\lambda_{sN}^{(3)}$ in (\ref{om4}). Thus in
the last line of (\ref{om4}) the eigenvalues $\lambda_{sN}^{(3)}$
corresponding to the eigenfunctions $p_n(x,y,z) =\cos(2\pi n z
/\beta)$ (for which $\lambda_{iN}=0$) are effectively subtracted.

Our strategy is the following: one expands the logarithms in the
formula (\ref{om4}) in series
\begin{equation}
\ln(1-\exp(-2a\lambda_i)) = -\sum_{n=1}^{+\infty} \frac{\exp(-2a n
\lambda_i)}{n}, \label{om7}
\end{equation}
applies the expansion (\ref{r22}) to each term in (\ref{om7}) for
short distances $a$ and performs the sum over $n$ thus getting
Riemann zeta function at integer positive values.

It is possible to obtain the coefficients of the heat kernel
expansion for the operators $\Delta^{(3)}$ along the following
lines. For the manifold $[0,\beta/2]\times M $ and Dirichlet
boundary conditions one can write:
\begin{multline}
\exp(-t\Delta^{(3)}) \underset{t\to 0}{\sim} \Biggl(
\frac{\beta/2}{\sqrt{4\pi}}\: t^{-1/2} - \frac{1}{2} \Biggr)  \\
\Biggl( \frac{S}{4\pi} t^{-1} - \frac{P}{4\sqrt{4\pi}} \: t^{-1/2} +
\sum_i \frac{1}{24} \Bigl(\frac{\pi}{\alpha_i}-
\frac{\alpha_i}{\pi}\Bigr) + \frac{1}{12\pi} \int_{\gamma_j}
L_{aa}(\gamma_j) dl \Biggr) \label{om6}
\end{multline}
Here $S$ is an area and $P$ is a perimeter of the cross section $M$,
$\alpha_i$ is the interior angle of each sharp corner at the
boundary $\partial M$ and $L_{aa} (\gamma_j)$ is the curvature of
each boundary smooth section described by the curve $\gamma_j$.

Thus the important for our purpose Seeley coefficients for the
operator $\Delta^{(3)}$ with Dirichlet boundary conditions imposed
at the boundary of the manifold $[0,\beta/2]\times M $  can be
obtained from (\ref{om6}) as the coefficients at specific powers
of $t$ in the expansion (\ref{om5}):
\begin{align}
c_{0D}^{(3)} &= \frac{\beta S}{2 (4\pi)^{3/2}}  \\
c_{1D}^{(3)} &=  - \frac{S}{8\pi} - \frac{\beta P}{32\pi} \\
c_{2D}^{(3)} &= \frac{\beta}{4\sqrt{\pi}} \Biggl( \sum_i
\frac{1}{24} \Bigl(\frac{\pi}{\alpha_i}- \frac{\alpha_i}{\pi}\Bigr)
+ \frac{1}{12\pi} \int_{\gamma_j} L_{aa}(\gamma_j) dl \Biggr)  +
\frac{P}{16\sqrt{\pi}}
\end{align}

One can check by a direct calculation that
$c_{0N}^{(3)}=c_{0D}^{(3)}$, $c_{1N}^{(3)}=-c_{1D}^{(3)}$,
$c_{2N}^{(3)}=c_{2D}^{(3)}$ for manifold $[0,\beta/2]\times M $
with Neumann boundary conditions.

The other needed coefficients for manifold $M$ can also be read
off from (\ref{om6}): $c_{1D}^{(2)} = - c_{1N}^{(2)} =
-P/(8\sqrt{\pi})$, for manifold $[0,\beta/2]$: $c_{0N}^{(1)}=
\beta/(4\sqrt{\pi})$.

For $a \ll \beta/(4\pi)$ and $\lambda_{1 D} a \ll 1$, $\lambda_{1 N}
a \ll 1$ one obtains from (\ref{r22}) and (\ref{om4}) the leading
terms for the free energy:
\begin{equation}
\mathcal{F}(a,\beta)|_{a \ll \beta/(4\pi), \: \lambda_{1 D} a \ll 1,
\: \lambda_{1 N} a \ll 1} = -\frac{\zeta_R(4)}{8\pi^2}\frac{S}{a^3}
- \frac{\zeta_R (2)}{4\pi a} (1 - 2 \chi) + O(1) , \label{r23}
\end{equation}
where
\begin{equation}
\chi = \sum_i \frac{1}{24} \Bigl(\frac{\pi}{\alpha_i}-
\frac{\alpha_i}{\pi}\Bigr) + \sum_j \frac{1}{12\pi}
\int_{\gamma_j} L_{aa} (\gamma_j) d\gamma_j.
\end{equation}
The force calculated from (\ref{r23}) coincides with the zero
temperature force $F_C$ in \cite{Jaffe2}, (Eq.7).  Thus we prove
that in the finite temperature case the leading short distance
terms are the same as in the zero temperature case.

In the limit $\lambda_{1 D} a \ll 1$, $\lambda_{1 N} a \ll 1$ one
immediately obtains from (\ref{r24}) the high temperature result for
two parallel perfectly conducting plates separated by a distance
$a$. One expands logarithms in series and uses (\ref{r22}) and
$c_{0D}=c_{0N}= S/(4\pi)$ in two dimensions ($d=2$) to obtain:
\begin{multline}
\mathcal{F}(a, \beta)|_{a \gg \beta/(4\pi),\: \lambda_{1 D} a \ll 1,
\: \lambda_{1 N} a \ll 1} =
\\ =  - \sum_{\lambda_{k D}} \frac{1}{2\beta} \sum_{n=1}^{+\infty}
\frac{\exp(-2an\lambda_{k D})}{n}\Bigl|_{a\to 0} - \sum_{\lambda_{i
N}} \frac{1}{2\beta} \sum_{n=1}^{+\infty} \frac{\exp(-2an\lambda_{i
N})}{n}\Bigr|_{a\to 0}
 = \\
= \sum_{n=1}^{+\infty} - \frac{1}{2\beta}
\frac{1}{n}\frac{1}{(2an)^2} 2 (c_{0D} + c_{0N}) =
-\frac{\zeta_R(3)}{\beta a^2} \frac{S}{8\pi}  , \label{om2}
\end{multline}
which is a well known result \cite{Brevik, Sauer, Lifshitz}.

\section*{Acknowledgments}

  V.M. thanks Dmitri Vassilevich
  for comments and suggestions on the paper.
  This work has been
  supported by  grants RNP $2.1.1.1112$ and
  SS $.5538.2006.2$.

\end{document}